# An Affordable Experimental Technique for SRAM Write Margin Characterization for Nanometer CMOS Technologies

**Bartomeu Alorda, Cristian Carmona, Gabriel Torrens, Sebastià Bota**

Electronic Systems Group, Physics Dept, Universitat Illes Balears, Spain

E-mail: tomeu.alorda@uib.es

*Abstract*— Increased process variability and reliability issues present a major challenge for future SRAM trends. Non-intrusive and accurate SRAM stability measurement is crucial for estimating yield in large SRAM arrays. Conventional SRAM variability metrics require including test structures that cannot be used to investigate cell bit fails in functional SRAM arrays. This work proposes the Word Line Voltage Margin (WLVM), defined as the maximum allowed word-line voltage drop during write operations, as a metric for the experimental characterization of write stability of SRAM cells. Their experimental measurement can be attained with minimal design modifications, while achieving good correlation with existing writability metrics. To demonstrate its feasibility, the distribution of WLVM values has been measured in an SRAM prototype implemented in 65 nm CMOS technology. The dependence of the metric with the width of the transistors has been also analysed, demonstrating their utility in post-process write stability characterization.

## I. INTRODUCTION

A significant percentage of the total die area of current System-on-Chip (SoC) is dedicated to memory blocks. One consequence of this fact is that embedded SRAM yield dominates the overall SoC yield [1]. The most common SRAM reliability issues are related to achieving correct cell writings and preventing accidental data changes when reading. Write and read stability margins have been defined to quantify the robustness related to a specific bit cell design against a possible failure during write or read operation respectively [2]. In six transistors (6T) bit cells, acceptable values for both write and read margins have been traditionally obtained by accurate transistor sizing, as a trade-off exist between them. In nanometer technologies, the reliability levels reached with the classical approach are affected by a noticeable reduction of write and read margins due to supply voltage reduction, and also, by the effects of process variability associated to technology scaling [2]. Variability on transistor parameters is directly transferred to variability on cell stability margins. Therefore, a deep knowledge and analysis about cell stability has become a must in the design of modern SRAM devices [3,4]. The identification of bit cells with stability margins affected by variability (weak cells), is essential to increase the overall memory reliability, as the probability to experience a failure during an access on those cells is significantly higher than in the rest. Although read stability usually has gathered more attention than write stability, an accurate estimation of writability is also becoming important in nanometer SRAMs, in this sense, Hirabayashi [5] reported that write operation failure is dominant in low power memories.

Cell stability characterization has been usually performed using electrical simulation. Monte Carlo analysis has been used to find the effect of process variations [6]; hence the obtained results depend on the accuracy of the available device models. In any case, an experimental characterization is indispensable, for example when identification of weak cells is required to prevent the use of low reliable memory addresses. In general, the main issue related to the experimental characterization of bit cell stability is due to loss of performance as a consequence of the design modifications required for the measurement of the stability margins.

This paper proposes a new technique for experimental characterization of cell writability. The method consist in measuring the Word-line Voltage Margin (WLVM) defined as the maximum word-line voltage reduction tolerated by the cell to successfully perform a write operation when both the latch cell and bit-line drivers are operated at nominal $V_{DD}$ [7]. An analogous parameter was proposed in [8] although they used bit-line current monitoring for their measurement. This work presents an alternative method that, instead of current probes, requires only a minimal row decoder redesign [9], without affecting memory timing and power consumption and with low impact on the decoder area. As a result, the proposed metric can provide a fast non-invasive characterization of writability. Furthermore, it can also be compatible with write-assist techniques [10].

## II. BIT-CELL WRITABILITY

A six transistor memory cell is formed by two access transistors controlled by the word-line (WL), and two back-to-back connected inverters forming a latch (Fig 1a). Cell ratio defines the ratio between the width of inverter pull-down devices and the access transistors, $CR=W_n/W_{tx}$; pull-up ratio defines the ratio between the width of inverter pull-up devices and the access transistors, $PR=W_p/W_{tx}$.

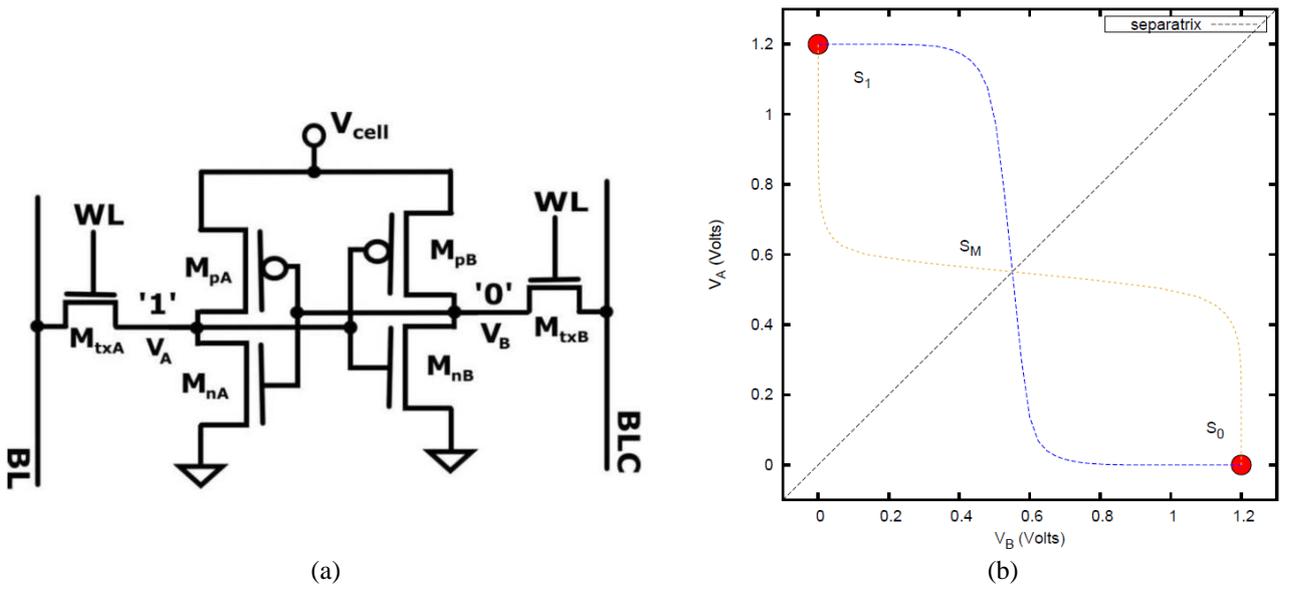

**Fig. 1**. (a) Six transistors SRAM cell, (b) voltage transfer characteristics.

The latch is a non-linear system where the voltage nodes $V_A$ and $V_B$ constitute the state vector $\mathbf{V} = (V_A, V_B)$ and current equations at these nodes constitute the state equations [11]. The state equations have two stable points $S_1$ and $S_0$, associated with logic states '1' and '0' respectively, and one metastable point $S_M$. In fact, these solutions are the intersection points of the back-to-back connected inverters voltage transfer curves (Fig. 1b). The metastable point corresponds to $\mathbf{V}_{trip} = (V_{trip}, V_{trip})$ which marks the boundary between logic 1 and logic 0 for nominal $V_{DD}$ under stationary conditions. $\mathbf{V}_{trip}$ is placed on the separatrix, defined as the boundary, which separates the two stability regions [12], the separatrix in this case is a straight line since the 6T cell is symmetrical.

Write operation in a bi-stable cell (like a 6T cell) can be described as a transition from one equilibrium state to the other. Assuming that the cell is initially at state $S_1$, to write state $S_0$ in the cell, the bit-line (BL) voltage is decreased to '0', the complementary bit-line (BLC) is biased to '1' and the access transistors $M_{txA}$ and $M_{txB}$ are driven to their ON mode. Node A being at logic '1' is connected to BL at logic '0' and node B at logic '0' to BLC at logic '1', as a result $V_A$ decreases and $V_B$ increases. If $V_A$ is pulled down below the separatrix, the inherent positive feedback of the cell comes into play, $M_{pB}$ becomes stronger than $M_{nB}$ and increases $V_B$ even more, which is also the gate voltage of $M_{nA}$, as a result, $M_{nA}$ starts to conduct and further pulls down $V_A$, eventually $V_B$ rises to $V_{DD}$ and $V_A$ falls to 0, thus flipping the logic states of the cell.

The state of the cell will change or not depending on transistor parameters and the magnitude and duration of the perturbation induced during the writing process. Fig. 2 shows the state space trajectories obtained from electrical simulation of cell writing under different perturbation scenarios (state $S_0$ has been written in a cell being initially at $S_1$). Curve (a) corresponds to a successful write operation under nominal conditions ($V_{cell}=V_{WL}=V_{BLC}=V_{DD}$). The rest of examples correspond to failed writing attempts (state $S_0$ was not reached after the write process): curve (b) shows the consequence of an insufficient perturbation due to a not enough high value of the word-line voltage; curve (c) occurs when the low-level BL voltage has not been reduced enough, curve (d) is a consequence of an insufficient write access time, curve (e) corresponds to a cell with an excessive $W_{pull-up}$ that prevents writing even under nominal conditions, and, finally curve (f) shows the same as above but reducing $V_{cell}$ by 10%.

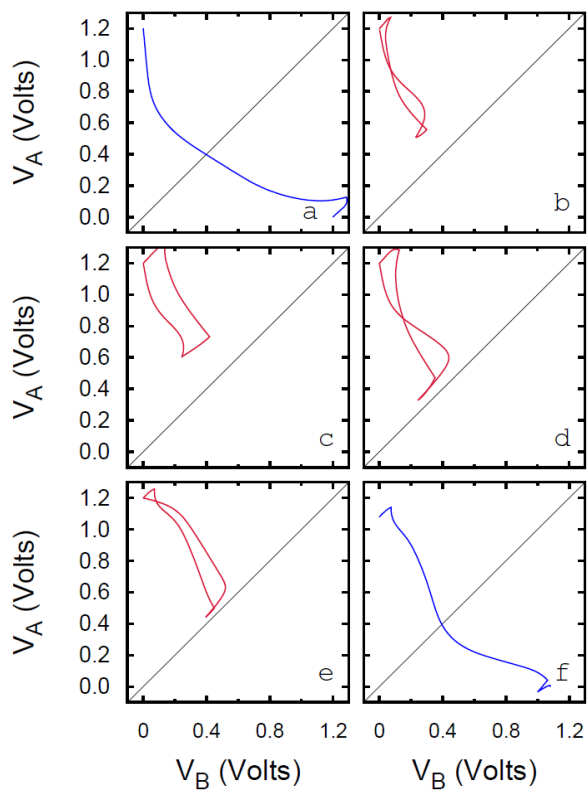

**Fig. 2.** State space trajectories during the write operation under different operating conditions. (a) Nominal condition, (b) reduced word-line voltage, (c) low-level bit-line voltage greater than 0, (d) Insufficient write access time, (e) excessive width of the pMOS transistor, (f) same as (e) with 10% reduction of $V_{cell}$.

When a memory is designed, it is intended that their bit cells should be written without difficulty even in the presence of process variations. Figure 2 shows that it should be useful to quantify the ability of a cell to be written. The term writability has been used to define this capability.

Traditionally, cell writability has been obtained computing the Write Noise Margin (WNM) [8]. The definition of the conventional WNM based on the butterfly curve is shown in Figure 3. The WNM is defined as the width of the smallest embedded square between the voltage transmission curves of inverter A (measured under $V_{WL} = V_{DD}$, $V_{BL} = 0$ V) and inverter B (measured under $V_{WL} = V_{DD}$, $V_{BLC} = V_{DD}$). A non-positive WNM indicates the impossibility of writing a new value. Despite its popularity, this parameter underestimates write cell capabilities due to its static nature [13].

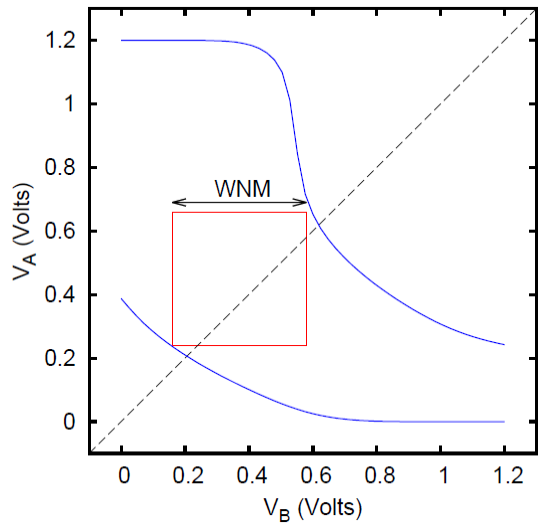

**Fig. 3.** Graphical definition of WNM.

As seen in Fig. 2, the dynamic stability of a bit cell is characterized by the transient state behaviour of the internal nodes during a write event, therefore, a definition based on a static behaviour does not seem the most appropriate. In this sense, there are alternative approaches that attempt to include the transient state behaviour of the internal nodes during a write event. These include the minimum (or critical) width of the word-line signal which causes the bit cell to flip its state [14], the maximum bit-line voltage able to flip the cell state or the minimum word-line voltage able to flip the cell during a write cycle [8].

### III. EXPERIMENTAL TECHNIQUES FOR WRITABILITY CHARACTERIZATION

Although easy to simulate, the experimental measurement of parameters like WNM, minimum word-line voltage or maximum bit-line voltage is difficult to perform since it requires access to internal nodes currents and voltages . In this sense, the different proposed techniques can be classified in three main categories according to their controllability and observability requirements: i) Techniques that require additional controllability and observability of internal cell nodes. ii) Techniques that require additional controllability and observability of common cell input/output signals like bit-lines and word-lines. iii) Techniques based on the control of timing parameters during access operations.

*A. Write Noise Margin*

Bhavnagarwala et al. [15] proposed an approach for the experimental measurement of WNM. They tried to replicate in real circuits the same technique used to measure WNM in electrical simulations. To do this, large analogue switch networks circuits were used to connect the two internal nodes of each memory cell to external sources and monitoring circuits. In practice, its application was limited to small memory arrays because the significant area overhead related to the analogue switch array and metal spacing constraints for routing out all internal nodes.

B. *Current-based techniques*

Other alternatives were based on measurements performed without a direct access to the internal cell nodes. In [8], word-line control and bit-line current monitoring has been proposed to characterize the stability parameters of large memory arrays without modifying the internal cell structure. Fig 4a shows a memory cell with initial state given by ($V_A$ = '0', $V_B$ = '1'), the cell supply node ($V_{cell}$), WL and BL are biased to $V_{DD}$, while the BLC node is ramped from $V_{DD}$ down to 0. The current measured on BL node ($I_{BL}$) is monitored vs. the BLC voltage, expecting a sudden drop in current as a consequence of a successfully write operation (Fig. 4b). The Bit-line Write Trip Voltage (BWTV) defined as the maximum BLC voltage, capable of flipping the cell in a write cycle, has been proposed [16-18] as a writability parameter.

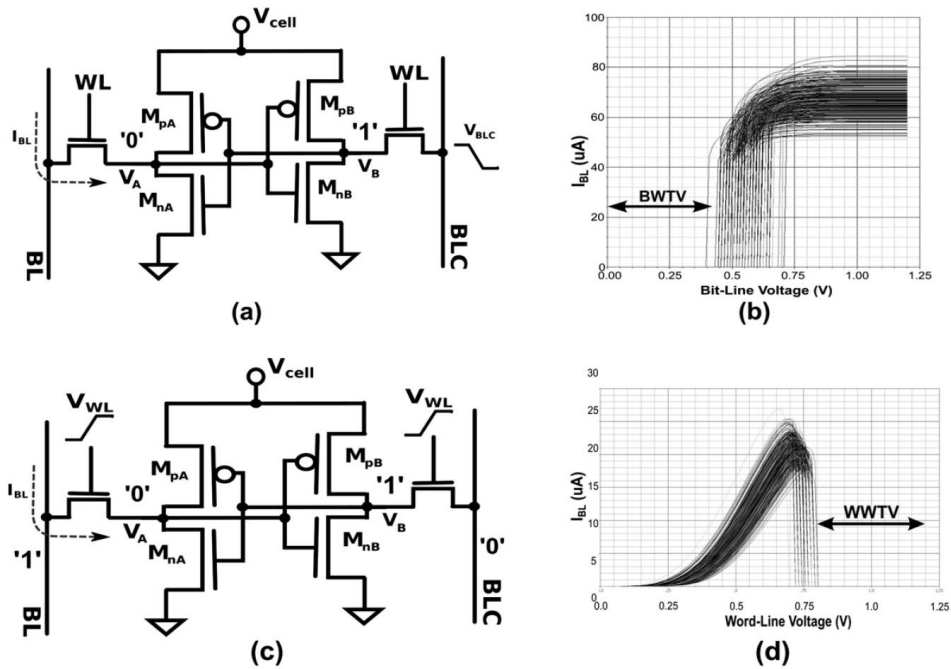

**Fig. 4.** (a) BWTV measurement scheme, (b) graphical definition of BWTV from Monte Carlo measured current curves using a commercial 65 nm CMOS technology, (c) WWTV measurement scheme, (d) graphical definition of WWTV from Monte Carlo measured current curves using a commercial 65 nm CMOS technology.

Alternatively, write margin has also been defined as the minimum WL voltage needed to perform a write operation [8]. The memory cell is written with an initial value while the bit-lines values are set to the complementary stored value (fig. 4c). The WL voltage is ramped up until a sudden transition of $I_{BL}$ is detected (fig. 4d), the current drops when the stored cell value has been changed.

The Word-line Write Trip Voltage (WWTV) was defined as the difference between the nominal power supply voltage and that minimum WL voltage needed to change the stored value.

The experimental measurement of both parameters only requires BL current monitoring and BL or WL control, therefore, an analogue switch array is still necessary to access all BL (although much simpler than for WNM measurement). Moreover, a new mode different from hold, read or write is also needed to test the device, and, finally, each memory cell has to be measured individually (a WL voltage ramp must be applied for each bit cell).

*C. Write margin estimation using read/write operations*

The most representative methodology of this group is based on finding the minimum WL pulse width to produce a successful write on a specific cell. In [14], the writability margin estimation based on word-line pulse timing control is implemented using a ring oscillator and frequency divider circuits. The memory instance is modified to add the pulse width modulation circuits to the word-line decoder, the rest of the memory blocks remain unaltered. This alternative reduces to a minimum level the memory array modification because only the decoder circuit should be modified; however, it faces the need to route the frequency divider signal output to all decoder outputs without introducing misbalanced delay between the WL pulse generation and the rest of memory control signals.

## IV. WORD LINE VOLTAGE MARGIN (WLVM)

WLVM is defined as the maximum reduction of the word-line voltage that still allows cell writing. Instead of monitoring the bit-line current as in the previous methods, its value can be measured applying a sequence of write operations of the complementary state of the stored value, performed at decreasing values of the word-line voltage. After each write operation of the sequence, the cell is read-out at nominal conditions to check if its state has changed or not.

First of all, the proposed metric has been checked by electrical simulation. Figure 5 shows simulation results for a commercial 65 nm CMOS technology obtained as a function of the pull-up ratio. The obtained WLVM values have been compared with the metrics presented in the previous section (WNM, BWTV, and WWTV). It is observed that WLVM presents a similar dependence with pull-up ratio than the others metrics. Note that WLVM is measured performing a sequence of standard write operations with reduced word-line voltages; unlike WWTV, measured while applying a continuous raising voltage ramp to WL. The fact that WWTV is measured when the cell current returns to zero at the end of the writing process, explains the lower value for WWTV observed in Fig. 5, moreover, it is interesting to note that, when measuring WWTV, the cell changes its state without experiencing the capacitive coupling effects appearing when pass transistors are turned off at the end of the write cycle.

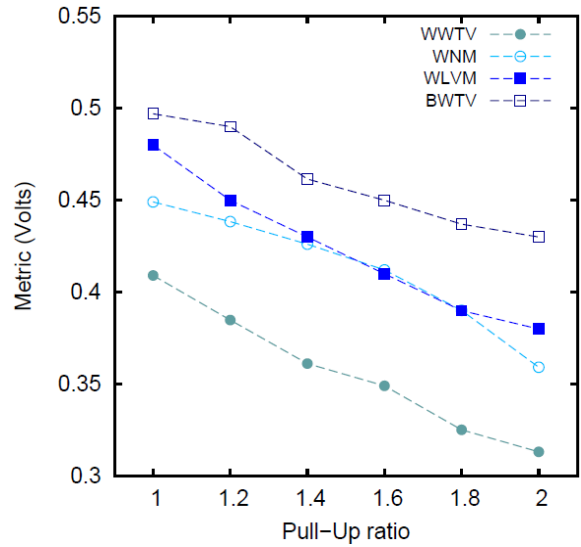

**Fig. 5.** Comparison of different writeability methods vs. PR (for CR=1).

Figure 6 shows the correlation obtained between BWTV, WWTV and WLVM metrics when process variation is taken into account. The quantization effects shown in Figure 6 are related to the $V_{WL}$ increments used in the simulations (the $V_{WL}$ was changed in steps of 10 mV). A linear correlation coefficient near 0.93 has been found in both cases. This result suggests the presence of a remarkable equivalence between the different metrics.

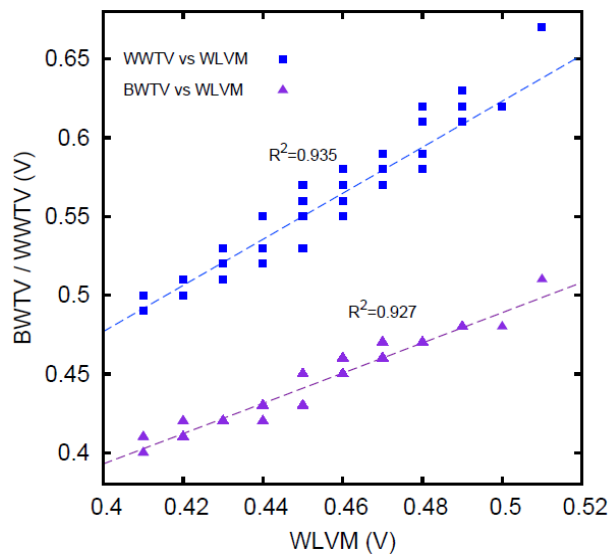

**Fig. 6.** Correlation between WWTV and BWTV with WLVM.

*A. Implementation Issues*

The experimental measurement of WLVM only requires controlling the word-line voltage during the write operation. This has been implemented by designing a row decoder with complemented outputs and adding an inverting stage with a dedicated power supply, $V_{DD\_WL}$ (Fig. 7). This last stage has been resized according to the word-line fan-in. Thereby, word-line voltage control is achieved without modifying the layout of the memory core.

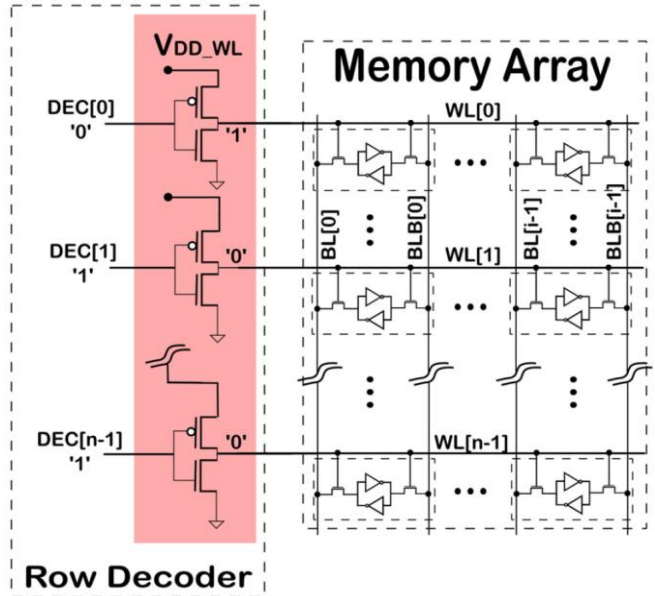

**Fig. 7.** Row decoder redesign for WLVM.

*B. WLVM Procedure*

Figure 8 shows the procedure followed for the experimental measurement of WLVM. First all memory cells are set to a known initial logic state, $S_X$ and a variable, *wlvm*, initiated to $V_{DD}$, is defined for each memory cell. Then, all cells are written to the complementary state $\overline{S_X}$ biasing $V_{DD\_WL}$ at $V_{DD}$. The logic state of all cells is checked with a sequence of read operations performed at nominal conditions, if the write operation failed in a cell with index *i*, their $wlvm_i$ value is updated to 0 ( *i* ranges from 0 to *N*, being *N* the number of memory cells). All write and read operations are performed sequentially byte by byte.

After recovering all the cells to $S_X$, a new write process to change cell state from $S_X$ to $\overline{S_X}$ is performed using a reduced value of the word-line voltage, $V_{DD\_WL} = V_{DD} - \Delta$. A subsequent read-out sequence is made under nominal conditions, the $wlvm_i$ values are updated to $\min(\Delta, wlvm_i)$ for those cells that have not changed their state to $\overline{S_X}$. The process is repeated decrementing the word-line voltage by Δ steps in successive iterations, $V_{DD\_WL} = V_{DD} - j \times \Delta$, and updating the $wlvm_i$ values to $\min(j \times \Delta, wlvm_i)$ in those cells that have failed writing the complementary state throughout iteration $j$. This procedure finishes when any cell can be written to their complementary value. Then, the process is repeated using $\overline{S_X}$ as initial value and trying to write $S_X$ in all memory cells. Throughout all the described sequence, all memory cells remain in hold mode for at least one clock cycle between any two consecutive accesses to the same or different cells.

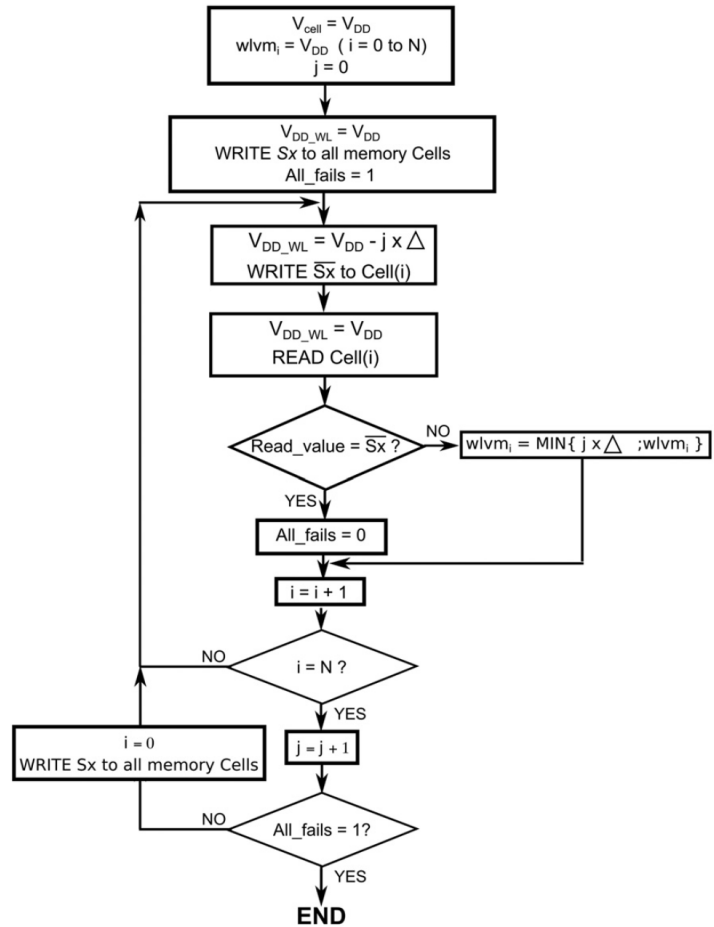

**Fig. 8.** Search algorithm for WLVM estimation. The algorithm must be applied for $S_x=0$ and then repeated for $S_x=1$.

Figure 9 represents the cumulative distribution function (*cdf*) and probability distribution function (*pdf*) of *wlvm* values for a memory device composed by 2048 memory cells obtained from Monte Carlo simulations considering parameter variations for a 65 nm CMOS commercial technology. Note that the weakest cell starts to fail at $wlvm_{min}$.

The WLVM technique can be applied at different levels (bit-cell, word, block…) depending on the object used to check for a successful write operation. At word level a *wlvm* variable is assigned to each word, at block level a variable is assigned to each block, only one variable is assigned at memory level, in all cases the test procedure is similar to the one described in fig. 8. In this way, it is possible to reduce test time in large devices. Note that, at memory level, the measurement ends when $wlvm_{min}$ has been found.

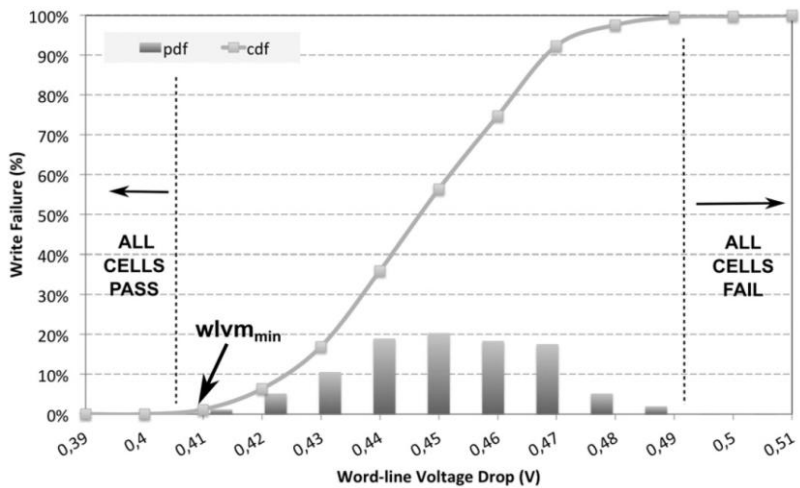

**Fig. 9.** Write Failure percentages obtained from Monte Carlo analysis.

## V. EXPERIMENTAL RESULTS

The WLVM technique has been validated on an experimental memory array implemented in a 65 nm CMOS commercial technology. The impact on area overhead related to the dedicated power supply node for word-line voltage modulation was observed to be negligible [10]. The implemented memory (fig. 10) contains five memory arrays of 256 rows and 16 columns (2048 cells) differing each one in their transistor sizing (table I). Bit cell layout was implemented following the design guidelines of regular-layouts to minimize the impact of variability. Such layout features includes using straight diffusion regions and regular alignment of word-line polysilicon lines without bends (CR was set to 1 for all cases) [19].

TABLE I. Transistor sizing.
All transistors have minimum channel length

| Cell | $M_n$ | $M_{tx}$ | $M_p$ | CR | PR |
|---|---|---|---|---|---|
| A | 150 nm | 150 nm | 150 nm | 1 | 1 |
| B | 150 nm | 150 nm | 230 nm | 1 | 1.5 |
| C | 150 nm | 150 nm | 300 nm | 1 | 2 |
| D | 230 nm | 230 nm | 230 nm | 1 | 1 |
| E | 300 nm | 300 nm | 150 nm | 1 | 0.5 |

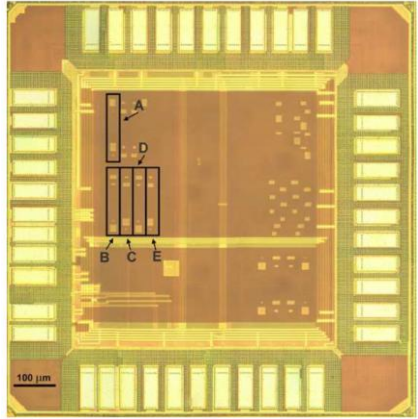

**Fig. 10.** Layout of the experimental SRAM prototype.

### A. Minimum-sized SRAM results

Memory A, having minimum sized transistors (PR=1 and CR=1), was checked at bit-cell level using the procedure described in Fig. 8. Figure 11 shows the resulting cumulative density function (*cdf*) and the probability density function (*pdf*) for the measured values of *wlvm*. Measurements were performed first with $S_0$ as initial state (curve labelled as *0to1*) and then with $S_1$ (*1to0*). We found that all cells were writeable and readable under nominal conditions. Due to variability, a difference of 160 mV is observed between the *wlvm* values of the strongest and weakest cells (address and index of the weakest cells were identified during the measurement). A small shift (~ 10 mV) between *0to1* and *1to0* curves is appreciated in the *cdf* (the same behaviour has been observed in all the measured samples). The origin of that difference was probably related to a small mismatch between BL and BLC drivers.

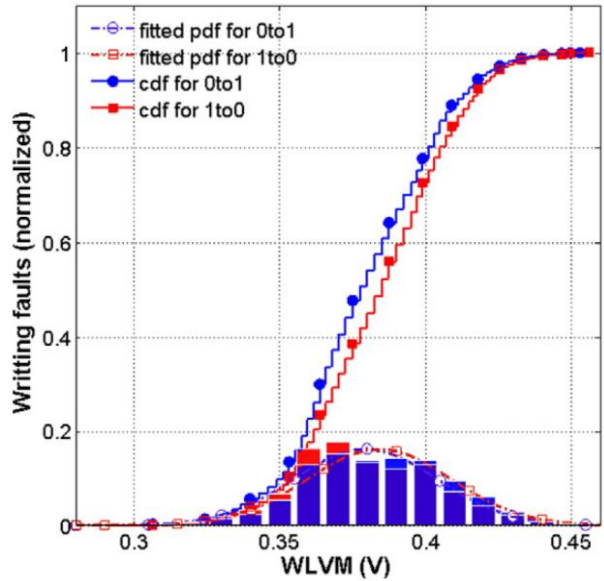

**Fig. 11.** Normalized pdf and cdf plots for the 0 to 1 (blue line) and 1 to 0 transitions (red line).

Figure 12 shows the scatter plot obtained with the pairs of values ($wlvm_{1to0}$, $wlvm_{0to1}$) of all memory cells. The value of the correlation coefficient between the two variables, $r=0.2826$, suggests a poor positive correlation between the two values (as $r^2=0.08$, this means that only 8% of the change in $wlvm_{1to0}$ can be explained by the change in $wlvm_{0to1}$). As $wlvm_{1to0}$ was mainly determined by $PR_A$, while $wlvm_{0to1}$ depends on $PR_B$ [2], Fig. 12 can be regarded as experimental evidence that the impact of variability is different on each cell transistor. Consequently, the identification of the cells having low write stability margins requires the measurement of both, $wlvm_{0to1}$ and $wlvm_{1to0}$. For each cell we define their $wlvm$ value as the minimum of $wlvm_{0to1}$ and $wlvm_{1to0}$.

The results for the technique applied at byte level are shown in fig. 13. In this case, only one $wlvm$ value is assigned for each group of eight bits, this value corresponds to the lower $wlvm$ cell value present in that byte. Consequently, both the $pdf$ spread and mean value have been reduced. Table II shows the parameters obtained in both cases assuming normal distributions.

TABLE II. WLVM parameters

|  | $WLVM_{min}$ (mV) | 0_to_1 transition (mV) | | 1_to_0 transition (mV) | |
| --- | --- | --- | --- | --- | --- |
|  |  | Mean | Standard deviation | Mean | Standard deviation |
| Bit level | 317.7 | 405.2 | 42.0 | 406.5 | 42.1 |
| Word level | 317.7 | 375.9 | 15.1 | 372.6 | 15.3 |

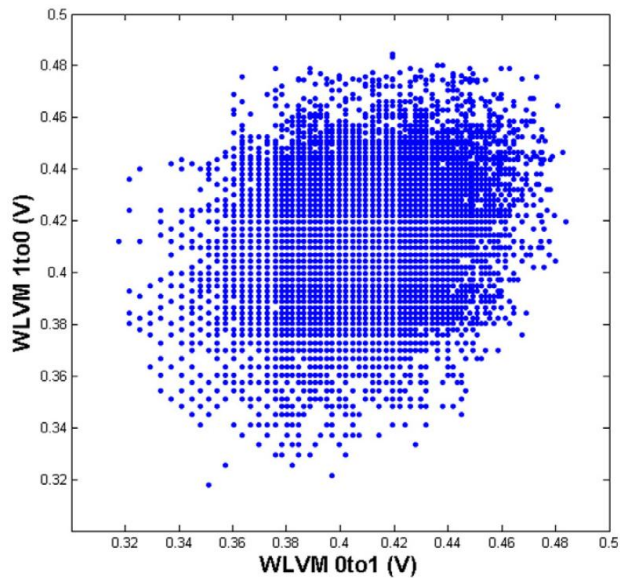

**Fig. 12.** Scatter plot of $wlvm_{1to0}$ vs. $wlvm_{0to1}$.

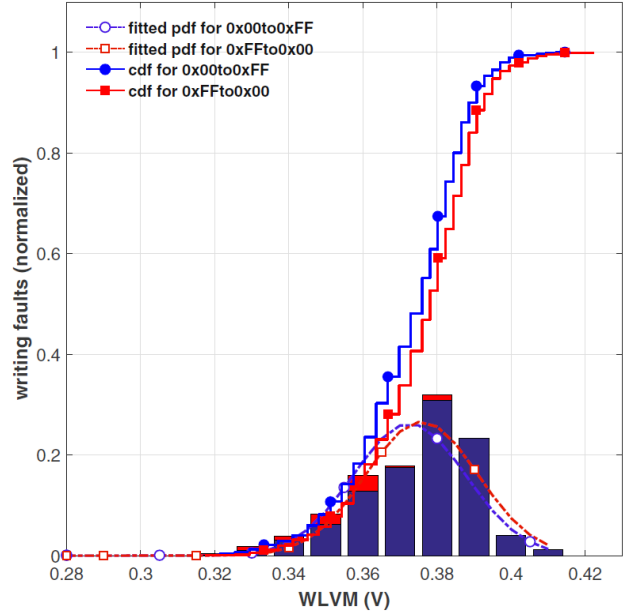

**Fig. 13.** Write Failure ratio on minimum-sized memory cells (cell A) corresponding to 0x00 to 0xFF and 0xFF to 0x00 transitions for each memory address.

### B. Dependence with Pull-up ratio

As it has been shown in Fig, 5, the ability of overwriting the cell content is enhanced when PR is reduced [2]. Minimum width pull-up and pass transistors are usually used to set the PR to a reasonable low value and minimise cell area; pull-down transistors are sized to attain a suitable read noise margin, CR value is usually chosen near 2, a good trade-off value to achieve enough stability without an excessive area increment. In some specific circumstances, it must be desirable further increase of PR, for instance, in [19] it was shown that SRAM soft error rate is decreased when PR is increased.

Figure 14 shows the *WLVM* cumulative distributions corresponding to the five cell designs reported in table I. Table III collects the mean value and standard deviation of the *pdfs* obtained from simulation and experimental measurements. The experimental results reasonably agree with the values obtained previously by electrical simulation, although only the experimental measurement can identify the exact location of the cells susceptible to cause write errors. As expected, a PR increment implied a shift toward lower voltages of the *cdfs* (less writeable cells). Writeability is enhanced when decreasing PR below 1, the mean value of *wlvm* improves by 17.5% if PR is reduced to 0.5, this involves using a minimum sized pull-up transistor and double the width of the pass and pull-down transistors, which, in our case results in an increase in cell area of 30%. Standard deviation was similar for cells with minimum size pass transistors (A, B and C); its value decreases ≈10% wherein the transistor width is greater (D and E).

We also note that *wlvm* not only depends on PR, the mean value of *wlvm* for cell D is 7.5% lower than for A, although both cells have the same PR and CR values. This result is also observed in the electrical simulations, the difference is probably related to the influence of short-channel effects on threshold voltage affecting the relative strength of the involved transistors. Fortunately the best result corresponds to the cell having minimum size transistors.

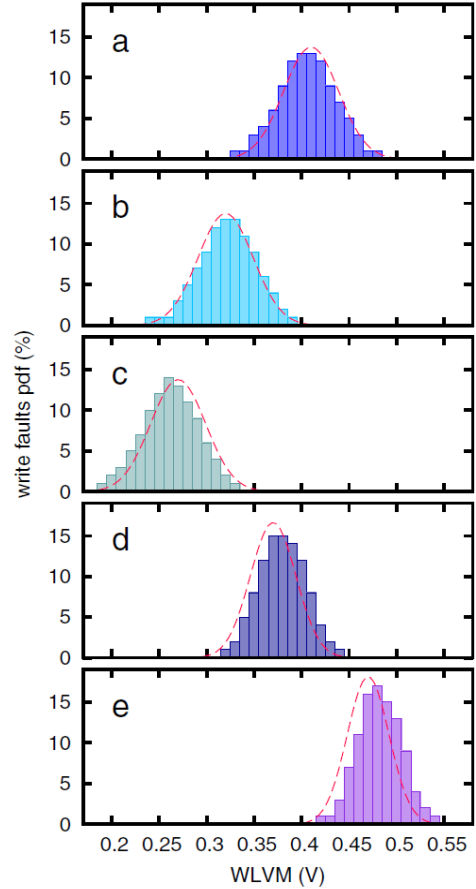

**Fig. 14.** Experimental WLVM distributions for the different cell types (dashed line corresponds to the result from Monte Carlo analysis).

TABLE III. Dependence on Pull-up ratio.

| Cell | PR | Simulated WLVM | | Experimental WLVM | |
|---|---|---|---|---|---|
| | | Mean (mV) | Std. dev. (mV) | mean (mV) | Std dev (mV) |
| A | 1 | 410 | 29 | 405 | 42 |
| B | 1.5 | 320 | 29 | 323 | 41 |
| C | 2 | 270 | 29 | 262 | 42 |
| D | 1 | 370 | 24 | 378 | 35 |
| E | 1/2 | 470 | 22 | 482 | 32 |

## VI. CONCLUSIONS

A new technique for the experimental characterization of bit cell writability based on the determination of the word line voltage margin, has been proposed and validated in SRAM devices implemented in a 65 nm CMOS technology. We have demonstrated that their practical implementation requires only the control of the word-line voltage as additional feature. In consequence, it is an affordable technique, as it does not require modifications of the memory core design (only the last stage of the row decoder has to be modified), resulting in a minor impact in timing, area and power consumption.

We observed, from electrical simulation, a good correlation of the proposed WLVM metric with previous metrics as WNM, BWTV or WMTV.

The application of the technique allows identification of non-writeable cells or of those cells that are harder to be overwritten. Indirectly, the technique can also be used to quantify the impact of variability in an electrical bit-cell characteristic like *wlvm*. Finally, we applied the technique to analyze the dependence of writability with Pull-up ratio.

### Acknowledgment

The Spanish Ministry of Science and Innovation has supported this work under project TEC2011-25017.

### References

[1] International technology roadmap for semiconductors. <http://www.itrs.net/> [October 2015].

[2] A. Pavlov, M. Sachdev, CMOS SRAM Circuit Design and Parametric Test in Nano-Scaled Technologies, Springer, NY 2008.

[3] J. Wang, S. Nalam, B.H. Calhoun, Analysing Static and Dynamic Write Margin for nanometers SRAMs, ACM/IEEE International Symposium on Low Power Electronics and Design (ISLPED) 2008, pp. 129-134.

[4] H. Aghababa, B. Ebrahimi, A. Afzali-Kusha, M. Pedram, Probability calculation of read failures in nano-scaled SRAM cells under process variations, Microelectron. Reliab. 52 (2012) 2805-2811.

[5] O. Hirabayashi, A. Kawasumi, A. Suzuki, Y. Takeyama, K. Kushiyama, T. Sasaki, A. Katayama, G. Fukano, Y. Fujimura, T. Nakazato, Y. Shizuki, N. Kushiyama, T. Yabe, A process-variation-tolerant dual power supply SRAM with 0.179 um$^2$ cell in 40nm CMOS using level-programmable wordline driver, IEEE International Solid-State Circuits Conference 2009, pp. 458-459.

[6] H. Makino, N. Okada, T. Matsumura, K. Nii, T. Yoshimura, S. Iwade, Y. Matsuda, Improved Evaluation Method for the SRAM Cell Write Margin by Word Line Voltage Acceleration, Circuits and Systems3 (2012) 242-251.

[7] C. Carmona, G. Torrens, B. Alorda, SRAM write margin cell estimation using wordline modulation and read/write operations, European Workshop on CMOS variability (VARI) 2014, pp 1-6.

[8] Z. Guo, A. Carlson, L. Pang, K. T. Duong, T. King Liu, B. Nikolic, Large-Scale SRAM Variability Characterization in 45nm CMOS, IEEE J. Solid-State Circuits 44 (2009) 3174-3192.

[9] B. Alorda C. Carmona, S. Bota. "Word-line power supply selector for stability improvement of embedded SRAMs in High Realiability Applications" Design Automation and Test in Europe Conference 2014, pp. 1-6.

[10] B. Alorda, G. Torrens, S. Bota, J. Segura, Adaptive static and dynamic noise margin improvement in minimum-sized 6T-SRAM cells, Microelectron. Reliab. 54 (2014) 2613-2620.

[11] B. Zhang, A. Arapostathis, S. Nassif, M. Orshansky, "Analytical modeling of SRAM dynamic stability" IEEE/ACM International Conference on Computer-Aided Design 2006, pp. 315–322.

[12] Y. Zhang,, P. Li, G.M. Huang, Separatrices in high-dimensional state space: system-theoretical tangent computation and application to SRAM dynamic stability analysis, 47th Design Automation Conference 2010, pp. 567-572.

[13] D. Khalil, M. Khellah, N. S. Kim, Y. Ismail, T. Karnik and V. K. De, Accurate Estimation of SRAM Dynamic Stability, IEEE Trans. Very Large Scale Integration (VLSI) Systems, 16 (2008) 1639-1647.

[14] Shao-Cheng Wang, Geng-Cing Lin, Yi-Wei Lin, Ming-Chien Tsai, Yi-Wei Chiu, Shyh-Jye Jou, Ching-Te Chuang, Nan-Chun Lien, Wei-Chiang Shih, Kuen-Di Lee, Jyun-Kai Chu, "Design and Implementation of Dynamic Word-Line Pulse Write Margin Monitor for SRAM" IEEE Asia Pacific Conference on Circuits and Systems 2012, pp. 116-119.

[15] A. Bhavnagarwala, S. Kosonocky, Yuen Chan, K. Stawiasz, U. Srinivasan, S. Kowalczyk, M. Ziegler, A sub-600 mV fluctuation tolerant 65nm CMOS SRAM array with dynamic cell biasing, IEEE J. Solid-State Circuits 43 (2008) 946-955.

[16] E. Grossar, M. Stucchi, K. Maex, W. Dehaene, Read stability and write-ability analysis of SRAM cells for nanometer technologies, IEEE J. Solid-State Circuits 41 (2006) 2577-2588.

[17] T. Fisher, E. Amirate, P. Huber, T. Nirschl, A. Olbrich, M. Ostermayr, D. Schmitt-Landsiedel, Analysis of read current and write trip voltage variability from a 1MB SRAM test structure, IEEE Trans. Semiconductor Manufacture 21 (2008) 534-541.

[18] N. Giercynski, B. Borot, N. Planes, H. Brut, "A new combined methodology for write-margin extraction of advanced SRAM", IEEE International Conference on Microelectronic Test Structures 2007, pp. 97-100.

[19] G. Torrens, S. A. Bota, B. Alorda, J. Segura, An experimental approach to accurate alpha-SER modeling and optimization through design parameters in 6T SRAM cells for deep-nanometer CMOS, IEEE Trans.Device. Mater. Reliab. 14 (2014) 1013-1021.